\documentstyle[aps,prl,twocolumn,epsf,float]{revtex}
\begin{document}

\twocolumn[\hsize\textwidth\columnwidth\hsize\csname @twocolumnfalse\endcsname

\title{
Topological Excitations of One-Dimensional Correlated Electron Systems
}

\author{
M.I.~Salkola$^{1,2}$ and J.R.~Schrieffer$^1$
}
\address{
$^1$NHMFL and Department of Physics, Florida State University, 
Tallahassee, Florida 32310 \\
$^2$Department of Physics, Stanford University, Stanford, California 94305
}

\date{August 11, 1998}
\maketitle

\begin{abstract}

Properties of low-energy excitations in one-dimensional superconductors and density-wave
systems are examined by the bosonization technique. In addition to the usual spin 
and charge quantum numbers, a new, independently measurable attribute is introduced 
to describe elementary, low-energy excitations. It can be defined as a number 
$w$ which determines, in multiple of $\pi$, how many times the phase of the 
order parameter winds as an excitation is transposed from far left to far right. 
The winding number is zero for electrons and holes with conventional quantum numbers, 
but it acquires a nontrivial value $w=1$ for neutral spin-${1\over 2}$ excitations
and for spinless excitations with a unit electron charge. It may even 
be irrational, if the charge is irrational. Thus, these excitations 
are topological, and they can be viewed as composite particles made of spin 
or charge degrees of freedom and dressed by kinks in the order parameter.

\vspace*{0.4truecm}


\end{abstract}

\vspace*{0.5truecm}

]


The concept of elementary excitations due to Landau plays a fundamental
role in understanding many properties of condensed-matter materials. 
It relies on the assumption that in the renormalization-group 
sense there exists a map onto an effective model with the same low-energy, 
long-wavelength physics but with few relevant degrees of freedom left.
The practical significance of this approach is that seemingly complex
systems can be studied and experimental predictions made
even though true microscopic interactions may be strong.

In spite of the success of the basic concept, it cannot be justified 
rigorously because of the break-down of the renormalizing procedure.  
Indeed, there are many interesting instances where the physical picture 
of weakly interacting elementary excitations with quantum numbers
equal to those of bare ones becomes invalid.
Examples of such cases are the quantum Hall effect \cite{pg} and quasi-one-dimensional
conductors \cite{hkss} where low-energy excitations have unusual spin-charge
relations and are not continuously related to conventional electrons and holes.

In this Note, we examine interacting one-dimensional conductors. This is
a widely relevant and frequently studied problem \cite{emery,solyom}. 
We consider a particular situation in which an extra particle is added 
into a superconductor. Yet, the consequences of this process illustrate that 
the elementary, low-energy excitations are more complex than one might 
have argued. Our most important result is that, in addition to the usual 
spin and charge quantum numbers, an excitation attaches itself a kink 
which can be quantified by another quantum number. It is defined as 
a number $w$ which determines, in multiple of $\pi$, how many times the phase
of the order parameter at 
a given point winds as the excitation is transposed from one to another 
end of the system. While the winding number $w$ does not constitute an 
independent degree of freedom in the sense that it would be unrelated
to the spin and the charge of the excitation, it does have physical 
attributes that make it measurable without any information about 
them. The winding number would be zero for an electron and a hole
with conventional quantum numbers, if they were to exist as elementary 
excitations, but it acquires a nontrivial value $w=1$ for a neutral 
spin-${1\over 2}$ excitation and for a spinless excitation with 
a unit electron or hole charge. It may even be irrational, if the charge 
is irrational. Thus, these excitations are topological objects which 
can be viewed as composite particles made of spin or charge degrees 
of freedom and dressed by $e^{i\pi w}$-phase kinks in the order parameter. 
The winding number appears naturally in systems which have continuous 
symmetry and whose ground states develop quasi-long-range order in 
one dimension and true long-range order in higher dimensions. 
In addition to superconductors, kinks must then be introduced in 
spin- and charge-density-wave systems. Our conclusion complements 
the earlier observation \cite{sol} that soliton excitations have 
unusual fermion quantum numbers in charge-density-wave systems 
where massless fermions are coupled to a boson field. Here we 
demonstrate that also the converse is true: elementary charge 
and spin excitations always carry kinks in one-dimensional systems 
with quasi-long-range order.

Specifically, consider the one-dimensional Hamiltonian
\begin{equation}
H=-t\sum_{n\sigma}(c^\dagger_{n+1\sigma}c_{n\sigma} + h.c.)
    + \mbox{$1\over2$} \sum_{nm} V_{n-m} \rho_{n}
 \rho_{m}, \label{eq:e1}
\end{equation}
where $c_{n\sigma}$ is the fermion operator for an electron of spin
$\sigma$ at site $n$, $\rho_{n} = \sum_\sigma c^\dagger_{n\sigma}c_{n\sigma}$ 
is the electron number density, 
$V_n$ is the electron-electron interaction, and $t$ is the nearest-neighbor
hopping matrix element.  
To address the low-energy and long-wavelength excitations, 
the energy spectrum is linearized at the Fermi energy and the fermionic degrees 
of freedom are expressed by slowly varying fields $\psi_{\sigma\pm}(x)$:
\begin{equation}
c_{n\sigma}/\sqrt{a} \sim \psi_{\sigma+}({x_n})e^{-ik_F x_n} + 
\psi_{\sigma-}(x_n)e^{ik_F x_n},
\end{equation}
where $k_F$ is the Fermi wavevector, $a$ is the lattice spacing, and 
$x_n=na$. Subsequently, the free part of the Hamiltonian, $H= H_0 + H_I$, 
becomes
\begin{equation}
H_0= v_F\int  \!\! dx\, [\psi^\dagger_{\sigma}({\bf x}_-) \hat{p}  
\psi_{\sigma}({\bf x}_-) -  \psi^\dagger_{\sigma}({\bf x}_+) \hat{p}  
\psi_{\sigma}({\bf x}_+)], \label{eq:free}
\end{equation}
where $\hat{p}= -i\hbar\partial_x$ is the momentum operator and
$v_F= (2ta/\hbar) \sin k_Fa$ is the Fermi velocity (hereafter a sum over
repeated indices is implied unless otherwise noted). The left ($+$) and 
right ($-$) moving electrons with spin $\sigma$ are labeled according 
to their arguments, denoting both the space and time coordinate, 
${\bf x}_\pm \equiv (x,t)_\pm$; namely,
$\psi_{\sigma}({\bf x}_\pm) = \psi_{\sigma\pm}({x,t})$. In the case of 
free fermions, the Heisenberg equations of motion, 
$i\hbar \partial_t \psi_{\sigma}({\bf x}_\pm) = [\psi_{\sigma}({\bf x}_\pm),H]$, 
lead to the field operators which are functions of variables $x \pm v_Ft$, 
where $\pm$ refer to the left- and right-moving electrons, 
respectively: $\psi_{\sigma}({\bf x}_\pm)= \psi_{\sigma}(x \pm v_Ft)$.

It is convenient to define left- and right-moving currents as
$J_{\sigma\sigma'}({\bf x}_\pm)= 
\hbox{:$\psi^\dagger_{\sigma}({\bf x}_\pm) \psi_{\sigma'}({\bf x}_\pm)$:}$. 
The colons denote normal ordering with respect 
to the filled Fermi sea of the noninteracting system. In terms of these currents,
the Hamiltonian describing the electron-electron interactions may be rewritten as
\begin{eqnarray}
& & H_I = \int \! \! dx \left(-V_1 J_{\sigma\mu}({\bf x}_+)J_{\mu\sigma}({\bf x}_-) 
  +  V_2 J_{\sigma\sigma}({\bf x}_+)J_{\mu\mu}({\bf x}_-)
\right. \nonumber\\
& & \  \ \  + \mbox{$1\over2$}V_3 [e^{i4k_Fx}
\hbox{:$\psi^\dagger_{\sigma}({\bf x}_-) \psi_{\sigma}({\bf x}_+)$:}\,
\hbox{:$\psi^\dagger_{\mu}({\bf x}_-)\psi_{\mu}({\bf x}_+)$:} + h.c.] \nonumber
\\
& & \ \ \  \left. +\mbox{$1\over2$}V_4 [J_{\sigma\sigma}({\bf x}_+) J_{\mu\mu}({\bf x}_+)
+J_{\sigma\sigma}({\bf x}_-)J_{\mu\mu}({\bf x}_-)]\right).\label{eq:int} 
\end{eqnarray}
As usual, $V_1$ is the backward-scattering constant, $V_2$ is the 
forward-scattering constant, $V_3$ is the Umklapp-scattering constant, 
and $V_4$ is another forward scattering constant. Because $V_3$ corresponds 
to the scattering processes which violate momentum conservation by $4k_F$, 
it is important only if $4k_F$ is equal to the reciprocal lattice constant 
(the second-order commensurability). Note that $V_{1,3}= V(2k_F)$ and 
$V_{2,4}=V(0)$, where $V(k)= \sum_n aV_n e^{ikna}$.
For simplicity, we consider a limit where the backward and Umklapp 
terms are either zero or scale to zero, so that the model is 
exactly solvable by bosonization. For example, far away from
the second-order commensurability, the Umklapp processes are effectively 
turned off and remain so under the renormalization-group flow. 
As long as the backward and Umklapp scattering terms remain irrelevant
in the sense of scaling, they can be neglected. Finally, the coupling 
constants $V_k$ are allowed to be independent, as implied by many 
important interactions that are not depicted by the original 
Hamiltonian, Eq.~(\ref{eq:e1}).

In Abelian bosonization \cite{emery,ml,abd}, the left- and right-moving 
fermion operators are expressed in terms of boson fields,
$\psi_{\sigma}({\bf x}_\pm) \sim K_{\sigma}
\hbox{:$\exp[\mp i \Phi_{\sigma}({\bf x}_\pm)]$:}$ (no sum over repeated 
indices is implied in this paragraph), 
where $K_{\sigma}$ is the Klein phase-operator which establishes 
the correct anticommutation relations for different fields  \cite{emery}.
For instance, with this definition, the current operator becomes 
$J_{\sigma\sigma}({\bf x}_\pm) =
(1/2\pi)\partial_{x} \Phi_{\sigma}({\bf x}_\pm)$. 
For $V_1= V_3=0$, the Hamiltonian
is quadratic in the boson fields $\Phi_{\sigma}$:
\begin{mathletters}
\begin{eqnarray}
H_0&=&  \sum_{\sigma}\hbar v_F \int {dx\over 4\pi}\left( [\partial_{x} \Phi_{\sigma}({\bf x}_+)]^2
+[\partial_{x} \Phi_{\sigma}({\bf x}_-)]^2\right), \\
H_I&=& \sum_{\sigma\mu} \int {dx\over 4\pi^2} 
\Big( V_2 [\partial_{x} \Phi_{\sigma}({\bf x}_+)][\partial_{x} 
\Phi_{\mu}({\bf x}_-)]  \\
& & \quad\quad \mbox{\ } +\mbox{$1\over2$}V_4\{[\partial_{x} 
\Phi_{\sigma}({\bf x}_+)]^2+[\partial_{x} \Phi_{\mu}({\bf x}_-)]^2\} \Big). \nonumber
\end{eqnarray}
In the absence of interactions, \end{mathletters}
$\Phi_{\sigma}({\bf x}_\pm)= \Phi_{\sigma}(x \pm v_Ft)$. $H$ is readily diagonalized by
a set of unitary transformations. First, define 
$\Phi_s({\bf x}_\pm)= [\Phi_\uparrow({\bf x}_\pm) - \Phi_\downarrow({\bf x}_\pm)]/\sqrt{2}$
and $\Phi_c({\bf x}_\pm)= [\Phi_\uparrow({\bf x}_\pm) + \Phi_\downarrow({\bf x}_\pm)]/\sqrt{2}$.
Second, define $\varphi_s({\bf x}_\pm)=\Phi_s({\bf x}_\pm)$ and
$
\varphi_c({\bf x}_\pm)= \Phi_c({\bf x}_\pm)\cosh\phi + 
\Phi_c({\bf x}_\mp)\sinh \phi,
$
with $g= V_2/(h v_F + V_4)$ and $\tanh 2\phi= g$. The Hamiltonian
is transformed into
\begin{equation}
H= \sum_{\nu=s,c}\hbar v_\nu \int {dx\over 4\pi} \left([\partial_{x}
 \varphi_{\nu}({\bf x}_+)]^2 +
[\partial_{x} \varphi_{\nu}({\bf x}_-)]^2\right),
\end{equation}
where $v_s= v_F$ and $v_c= v_F/\cosh 2\theta$ are the spin and charge velocities.
In the Heisenberg picture, $\varphi_s({\bf x}_\pm)= \varphi_s(x \pm v_st)$ and
$\varphi_c({\bf x}_\pm)= \varphi_c(x \pm v_ct)$. A useful relation in
computing correlation functions of interacting fields is the two-point
correlation, $\langle \varphi_\nu({\bf x}_\pm) \varphi_\nu(0_\pm)\rangle
= - \log (a\mp ix_{\nu\pm})$, where $x_{\nu\pm}= x \pm v_\nu t$ ($\nu =s,c$). 
The ultraviolet cutoff, the Fermi length, is formally associated with the 
lattice spacing $a$.

For $g < 0$, the superconducting (SC) instability is the most dominant 
one at zero temperature. Defining the singlet pairing field as
\begin{equation}
\Delta({\bf x})= [\psi_{\downarrow}({\bf x}_+)\psi_{\uparrow}({\bf x}_-) 
 - \psi_{\uparrow}({\bf x}_+)\psi_{\downarrow}({\bf x}_-)]/\sqrt{2},
\end{equation}
the correlation function
\begin{equation}
G^{(0)}_{SC}({\bf x}-{\bf y})= \langle 0| \Delta^\dagger ({\bf x}) \Delta({\bf y})|0 \rangle
\end{equation}
probes the ordering fluctuations of interest in the ground state $|0\rangle$. 
Below, we 
will always consider cases where ${\bf x}$ and ${\bf y}$ are measured
at equal times $t$: ${\bf x}= (x,t)$ and ${\bf y}= (y,t)$.  Using the identity,
\begin{equation}
:\!e^{i\alpha\varphi({\bf x})}\!:\, :\!e^{i\beta\varphi({\bf y})}\!:\,=\, 
:\!e^{i\alpha\varphi({\bf x})+i\beta\varphi({\bf x})}\!:
e^{-\alpha\beta\langle \varphi({\bf x})\varphi({\bf y})\rangle},
\end{equation}
one can show that
\begin{equation}
\Delta^\dagger({\bf x})\Delta({\bf y}) = G_{SC}(x-y) :\!e^{i\eta A_c}\cos A_s\!\!: , \label{eq:scc}
\end{equation}
where the prefactor is
\begin{equation}
G_{SC}(r)= \left({1\over a}\right)^\mu
\left({a^2 \over a^2 + r^2}\right)^{\mu_{SC}/2}.
\end{equation}
The operators, 
$A_\nu = [\varphi_\nu({\bf x})-\varphi_\nu({\bf y})]/\sqrt{2}$ ($\nu=s,c$),
are expressed in terms of new fields, 
$\varphi_s({\bf x})= \varphi_s({\bf x}_+) + \varphi_s({\bf x}_-)$ and
$\varphi_c({\bf x})= \varphi_c({\bf x}_+) - \varphi_c({\bf x}_-)$.
The exponents are defined as 
$\mu_{SC}= 1 + \eta^2$, $\mu= 1 + 1/\sqrt{1-g^2}$, 
and $\eta= [(1+g)/(1-g)]^{1/4}$.
In the ground state, $G^{(0)}_{SC}({\bf x})= G_{SC}(x)$ --- thus,
the asymptotic forms of the correlation function are
$G^{(0)}_{SC}({\bf x}) \propto x^{-\mu_{SC}}$ and $G^{(0)}_{SC}(q=0,\omega) 
\propto \omega^{-\alpha_{SC}}$, where $\alpha_{SC} = 2 - \mu_{SC}$. 
The superconducting correlation function is singular ($\alpha_{SC} > 0$) 
at low energies and small momenta, for $g<0$.

The effect of excitations on pairing fluctuations can be examined by
computing the correlation function in the state
$|\Psi_\uparrow\rangle = \psi_\uparrow^\dagger(0)|0\rangle$, where
$\psi^\dagger_\uparrow(0)= [\psi^\dagger_\uparrow(0_+) + 
\psi^\dagger_\uparrow(0_-)]/\sqrt{2}$ adds an electron into the system at the origin and
time $t=0$.
First, let the electron be far away from the points $x$ and $y$ where
the pairing fields are measured, $|r| \ll |R|$; here, $r=x-y$ and $R=(x+y)/2$. 
Then, writing $A_\nu=(r/\sqrt{2})\partial_x\varphi_\nu(R) + {\cal O}(r^3)$,
an operator-product expansion can be developed for Eq.~(\ref{eq:scc}).
The correlation function, 
\begin{equation}
G_{SC}^{(1)}({\bf x},{\bf y} )=  \langle\Psi_\uparrow|\Delta^\dagger ({\bf x}) \Delta({\bf y})| 
\Psi_\uparrow\rangle,
\end{equation}
simply becomes for $t=0$
\begin{equation}
G_{SC}^{(1)}({\bf x},{\bf y} )= \left({1\over a}\right)^{3\mu/2}
\left({a^2 \over a^2 + r^2}\right)^{\mu_{SC}/2} 
   \cos^2 {ar/2\over a^2 + R^2}.
 \label{eq:small}
\end{equation}
For given $r$, the influence of the electron on the correlation
function decays as $R^{-4}$. This behavior is universal as the scaling exponent 
is independent of the strength of interactions. The result is consistent with 
the natural observation that a single electron has no effect on the bulk 
properties of a superconductor.

Second, let $x$ and $y$ be arbitrary. The correlation function can be
written in the form
\begin{equation}
G_{SC}^{(1)}({\bf x},{\bf y} )= G_{SC}(r )F({\bf x},{\bf y}),
\end{equation}
where $F({\bf x},{\bf y})= [F_+({\bf x},{\bf y}) +
 F_-({\bf x},{\bf y})]/2$ and
\begin{eqnarray}
F_\pm({\bf x},{\bf y})&=& \cos [\theta_a(x_{s\pm})-\theta_a(y_{s\pm})] \\
& & \times e^{\pm iw_1[\theta_a(x_{c\pm})-\theta_a(y_{c\pm})]\pm iw_2[\theta_a(x_{c\mp})-\theta_a(y_{c\mp})]}.\nonumber
\end{eqnarray}
We have defined $\theta_a(x)= \arctan(x/ a)$, $w_1=\eta\cosh\phi$, and 
$w_2=-\eta\sinh\phi$ ($w_1 > w_2> 0$, for $g<0$). Note that
$\theta_a(x)\rightarrow {\pi \over 2}{\rm sgn}(x)$, as $|x|/a\rightarrow \infty$. 
Initially ($t = 0$), the function $F$ reduces to 
$F({\bf x},{\bf y})= \cos^2 [\theta_a(x)-\theta_a(y)]$. For $|r| \ll |R|$, we recover
Eq.~(\ref{eq:small}).
In the scaling limit where $r,R\rightarrow\infty$ with 
$z \equiv r/R < 1$ fixed, the asymptotic behavior of $G_{SC}^{(1)}$ is
\begin{equation}
G_{SC}^{(1)}({\bf x},{\bf y} ) \propto
\left({1\over R}\right)^{\mu_{SC}} \left[1 - \left({\lambda\over R}\right)^2\right],
\end{equation}
where $\lambda/a= [2z/(1-z^2)]^2$. Again, the scaling exponent describing 
the suppression of superconducting correlations near the electron is 
universal.

\begin{floating}[t] 
\vskip -.3truecm\narrowtext
\begin{figure}
\epsfxsize=7.5truecm
\hspace*{4truemm} \epsffile{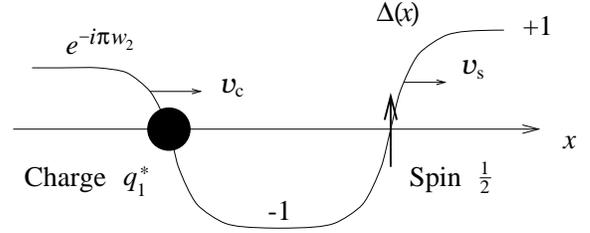}

\

\caption[]{\small
\noindent Schematic illustration of propagating kinks in the order parameter 
$\Delta({\bf x})\propto F({\bf x},\infty)$ after a right-moving
electron has broken up into spin and charge excitations in a superconductor 
away from half filling; $V_1=0$ and $V_2< 0$. The left-moving charge 
component is not shown. Note that, in the interacting 
system, the spin velocity is larger than the charge velocity, $v_s > v_c$.
}
\end{figure}
\vskip -.5truecm \end{floating}

In general, the order parameter $\Delta$ is obtained from 
$\langle\Delta^\dagger({\bf x})\Delta({\bf y})\rangle \rightarrow |\Delta|^2$, 
as $r\rightarrow \infty$. Likewise, the influence of an excitation on 
the superconductivity can be studied by letting $x$ to probe the order parameter 
in the neighborhood of the excitation and $y\rightarrow \infty$, even though 
strictly speaking $\Delta$ vanishes in one dimension.
For $t=0$, the solution is
\begin{equation}
F({\bf x},\infty) = 1 - {a^2 \over a^2 + x^2},
\end{equation}
which shows that the electron destroys the order parameter in 
its vicinity \cite{fl} while preserving $F\ge 0$. The order-parameter 
suppression dies out as $x^{-2}$. However, in the interacting system, 
the electron decays instantaneously into elementary excitations involving 
either spin or charge degrees of freedom. The kinky nature of these excitations 
is revealed, when the system is allowed to evolve in time. For illustrative 
purposes, consider the ground state with one right-moving electron initially 
added at the origin, $|\Psi_\uparrow\rangle = \psi_\uparrow^\dagger(0_-)|0\rangle$, 
so that $F= F_-$. The spin and the charge excitations carry kinks 
in the order parameter, and they both separately contribute to the total
phase shift an equal amount of $\pi$, because $w_1 + w_2 = 1$. Initially, 
the kinks overlap but then split as 
the excitations move apart. If only the spin excitation is located between 
the points $x$ and $y$ at a later time $t$, the correlation function is 
negative: $G_{SC}^{(1)}$ changes sign if either $x$ or $y$ crosses the position 
of the spin. This clearly shows that the state $|\Psi_\uparrow\rangle$ has 
an ordinary kink, or an antiphase domain wall, in the order parameter 
precisely at the position of the spin \cite{ts}.  The winding number of 
a spin-${1\over 2}$ particle is 1. In contrast, the splitting of the charge 
into right and left moving charge components, $q^\ast_1$ and $q^\ast_2$, 
leads to two irrational kinks whose winding numbers are $w_1$ and $w_2$.
The winding number of the kink is uniquely related to its charge ---
in units where the electron charge is one, the relation is particularly
simple: $w_{1,2} = q^\ast_{1,2}$. The total winding number, as well as 
the total charge, is conserved. Because the spin and charge velocities 
differ, the spin and charge kinks also propagate with the different 
velocities. This suggests a new and potentially attractive way to detect
spin-charge separation in a superconductor by using the Josephson effect.
It would also serve as evidence of irrational charge. Figure 1 describes 
schematically the effect of spin and charge on the superconducting 
order parameter.

The above result is generalized in a straightforward manner for $N$ electrons initially 
injected into the system at positions ${\bf R}_i=(R_i,0)$ ($i=1,\ldots,N$):
\begin{equation}
G_{SC}^{(N)}({\bf x},{\bf y} )= G_{SC}(r )\prod_{i=1}^N F({\bf x}-{\bf R}_i,{\bf y}-{\bf R}_i).
\end{equation}
If an infinite number of electrons is randomly distributed with 
a mean distance $\ell$, the correlation function decays exponentially:
\begin{equation}
G_{SC}^{(\infty)}({\bf x},{\bf y} ) \propto \left({1\over r}\right)^{\mu_{SC}} e^{-4r(t)/\ell},
\end{equation}
with $r(t)= \min [|r|,(v_s-v_c)|t|]$.
Thus, for any nonzero concentration of injected electrons, $n_{ex}=a/\ell \ne 0$,
superconductivity is destroyed.

One may also ask whether other quantities than the order parameter
show any signatures of kinks. Indirect evidence of kinks could be looked
for in the density of states and the conductivity, for example.
They are probed in specific heat and optical absorption measurements.
($i$) That there are new states at low energies (``midgap'' states) 
associated with the kinks is evident from the density of states 
$N(\omega)$. In the absence of excitations, at zero temperature, 
$N(\omega) \sim \omega^\alpha$, where $\alpha={1\over 2}(1/\sqrt{1-g^2}-1)$. 
Enhanced superconducting fluctuations will always lead 
to a pseudo-gap in the density of states, because $\alpha > 0$, for $g <0$. 
If there are excitations present in the system,
the density of states must correspondingly be modified at low energies such that
\begin{equation}
N(\omega) \propto (\omega^2 + \gamma_1^2)^{\alpha/2} + (\omega^2 + \gamma_2^2)^{\alpha/2},
\end{equation}
where $\gamma_1=(2v_c/\ell)[\cosh^2\phi-\sin({\pi\over 2}\cosh 2\phi)]$ 
and $\gamma_2=(2v_c/\ell)[1-\cos({\pi\over 2}\sinh 2\phi)]$. 
Thus, nonzero $N(\omega\rightarrow 0)$ implies that kinks give rise
to new states at low energies.
($ii$) The conductivity, however, does not exhibit any unusual behavior due 
to kinks, because it is sensitive to the concentration
of charges and the interactions between them --- injected electrons behave
undistinguishably from the existing particles, and the charge dynamics 
remains the same. If kinks and excitations were pinned, different kind 
of behavior would be expected.

In general, a nonzero backward scattering term which scatters electrons 
of opposite spin across the Fermi surface in opposite directions must 
be included. As a result, left- and right-moving spin degrees of freedom 
are coupled so that the spin will also split into left- and 
right-moving components. While it appears that the spin by itself cannot
have other values than integers and half integers, irrational values 
are allowed for the average spin and the concomitant kink.

\begin{mathletters}
The concept of a winding number applies equally to charge-density-wave
and spin-density-wave instabilities (at $2k_F$) that are described by operators
\begin{equation}
\rho({\bf x}) = \psi^\dagger_\sigma({\bf x}_+)\psi_\sigma({\bf x}_-)e^{i2k_Fx} + h.c.
\end{equation}
(charge-density-wave) and
\begin{equation}
S_a({\bf x}) = \psi^\dagger_\sigma({\bf x}_+)\tau^a_{\sigma\sigma'}\
psi_{\sigma'}({\bf x}_-)e^{i2k_Fx} + h.c.
\end{equation}
(spin-density-wave); $\tau^a$ are the three Pauli matrices ($a=1,2,3$).
The only difference is that the functional forms of the irrational winding 
numbers $w$ depend in a unique fashion on the nature of quasi-long-range order.
\end{mathletters}

In conclusion, elementary excitations in interacting one-dimensional conductors 
always carry kinks in the order parameter. The winding number characterizing
a kink cannot be arbitrary but is determined by the spin and the charge
of the excitation and by quasi-long-range order of the ground state --- 
in other words, by the interactions. This is not an entirely unexpected 
result, because electrons are interpreted in bosonization as solitons 
\cite{mand}. The novelty of our formulation is that the kink structure
of the excitations becomes observable, if the ground state develops 
quasi-long-range order. This implies a completely new conception 
of probing unusual quantum numbers of elementary excitations, 
which manifest themselves through structural and dynamical modulations
in the order parameter. 
For example, in a superconductor, the Josephson effect could provide
a method to measure winding numbers associated with injected elementary 
excitations. Using Hartree-Fock theory, we have already pointed out that 
spin excitations in superconductors tend to form antiphase domain walls 
(kinks) in the order parameter \cite{ss}. An analogous observation 
concerning impurity spinons has been made in the context of 
one-dimensional Kondo lattices \cite{zke}. Thus, these results 
corroborate the conclusion that kinks are generic excitations of 
superconductors.

We are grateful to Steven Kivelson for his 
insightful comments on the manuscript. M.I.S.~is indebted to Alexander
Fetter for his hospitality at Stanford University, where this work 
was completed. The support by the NSF under Grant 
Nos.\ DMR-9527035 and DMR-9629987 and by the U.S.~Department of 
Energy under Grant No.\ DE-FG05-94ER45518 is gratefully acknowledged.


\begin{references}
\bibitem{fl} Obviously, the exact form of the short-distance behavior 
is sensitive to the details of the ultraviolet cutoff, although
the relevant length scale is still the Fermi length $a$. Also,
fast oscillations at $2k_F$ have been neglected.
\bibitem{ts} In a triplet superconductor, on the other hand, different
forms of $F_-$ correspond to different values of the spin projection.
The spin excitation has the same kink profile in the $S_z=0$ component 
of the triplet order parameter as in the singlet order parameter, 
while the $S_z=\pm 1$ components have pure phase twists through complex 
values (cf. charge kinks).
\end{references}
\end{document}